\documentclass{article}

\usepackage[utf8]{inputenc}
\usepackage{graphicx}
\usepackage{color, colortbl}
\usepackage{todonotes}
\usepackage{amsmath}
\usepackage{amsthm}
\usepackage{listings}
\usepackage{xspace}
\usepackage{arxiv}

\newtheorem{definition}{Definition}
\newcommand*{\pfara}{{\sc PFaRA}\xspace}

\title{{\sc PFaRA}: A Platoon Forming and Routing Algorithm for Same-day Deliveries}

\author{
  S\^{\i}nziana-Maria Sebe\\
  orcidID: 0000-0002-9435-9879\\
  Institute of Informatics\\
  Clausthal University of Technology\\
  Clausthal-Zellerfeld, Germany\\
  \texttt{sinziana-maria.sebe@tu-clausthal.de} \\
   \And
 Jörg P. Müller \\
  orcidID: 0000-0001-7533-3852\\
  Institute of Informatics\\
  Clausthal University of Technology\\
  Clausthal-Zellerfeld, Germany\\
  \texttt{joerg.mueller@tu-clausthal.de} \\
}

\date{}

\begin{document}

\maketitle 

\begin{abstract}
Platoons, vehicles that travel very close together acting as one, promise to improve road usage on freeways and city roads alike. We study platoon formation in the context of same-day delivery in urban environments. Multiple self-interested logistic service providers (LSP) carry out same-day deliveries by deploying autonomous electric vehicles that are capable of forming and traveling in platoons. The novel aspect that we consider in our research is heterogeneity of platoons in the sense that vehicles are equipped with different capabilities and constraints, and belong to different providers. Our aim is to examine how these platoons can form and their potential properties and benefits. We present a platoon forming and routing algorithm, called \pfara, that finds longest common routes for multiple vehicles, while also respecting vehicle preferences and constraints. \pfara consists of two parts, a speed clustering step and a linear optimisation step. To test the approach, a simulation was used, working with realistic urban network data and background traffic models. Our results showed that the performance of our approach is comparable to a simple route-matching one, but it leads to better utility values for vehicles and by extension the LSPs. We show that the grouping provided is viable and provides benefits to all vehicles participating in the platoon.
\end{abstract}

\section{Introduction}
Platooning is a topical subject researched by many scientist especially in the context of automated driving. Relying on a leader to make the driving decisions and then mirroring them is one part of how this automation can be achieved. 
Research on platooning has been conducted since the 1950's \cite{lewis1958platoon}, \cite{schuhl1955probability} and has been gaining momentum since. Most platooning research tends to focus on highway scenarios \cite{swaroop1999constant} with large freight transports in mind e.g.~SARTRE \cite{robinson2010operating}. However, with recent developments in technology, urban platooning appears promising with respect to the better utilisation of the scare resource of urban space. However, current research on urban platooning focuses on stability, manoeuvring, and control (see e.g., \cite{ploeg2013lp}\cite{schindler2018dynamic} \cite{ali2015urban}), and, to a lesser extent, on the traffic management perspective (\cite{lioris2017platoons}, \cite{haas2018autonomous}). 

Same-day deliveries in the scope of logistic traffic is a novel use case for platooning. Due to the growth of online shopping, and the desire of customers for immediate shipping, logistic companies have to restructure the way they carry out their deliveries. \textit{On the one hand}, the distribution network has to be adapted. To support such orders, Crainic et al. developed a two-tier network architecture(\cite{crainic2008city}, \cite{crainic2014service}), with hubs, satellites and customers being the main transitional points. Goods are first carried from a large hub, located mostly outside the urban area, to multiple smaller locations called satellites, which are scattered inside the city, each catering to about 20 to 25 customers \cite{crainic2010two}. From the satellite, the goods are transported to the customer, the so-called last mile delivery, which accounts for the majority of the costs of said delivery \cite{gevaers2011characteristics}.

\textit{On the other hand}, the way orders are handled has to change. Usually, logistic companies plan and schedule deliveries in a way that minimises cost and maximises order completion. This process is costly in time and computational resources, making it an unviable approach for same-day orders. Immediate deployment of received orders, combined with having small autonomous electric vehicles carry out the costly last-mile stretch would be a possible solution. These vehicles are flexible and do not contribute as much to traffic congestion and toxic emissions, but sending multiple such vehicles in an already busy network does not alleviate the situation entirely.

Combining the logistic same-day deliveries into platoons could benefit urban traffic by streamlining deliveries, removing large trucks from the inner city and potentially decongesting some intersections. However, this subject has not been thoroughly researched thus far, with only a few publications existing (e.g ~\cite{scherr2018service},\cite{haas2018autonomous}) and to our knowledge, none that looks into how vehicles are grouped into platoons, especially with vehicles belonging to different service providers. We refer to Section \ref{sec:sota} for a more in depth analysis of the state of the art.

For our research we assume the context of logistic traffic, with multiple service providers, each with their own fleet of vehicles that carry out deliveries to customers. We also assume that platooning is incentivised by the traffic management authority, providing platooning vehicles with subsidies. Platoons form in a spontaneous way, at intersections where more than two vehicles meet.  A vehicle only forms/joins a platoon if its perceived utility for joining is better than if it were to travel alone. In order to gain the most benefit from platoons, we assume that logistic service providers are willing to form mixed platoons with each other. This leads to a high degree of heterogeneity, not only in the vehicles and destinations sense, but given the self-interested nature of all providers; in valuation functions, restrictions and preferences as well. This research presents a grouping and routing algorithm that fosters this type of cooperation while respecting the limitations imposed.

Our previous work \cite{vehits19} presented a first and simple optimisation approach to solve the problem of cross-provider platoon formation. In this work the optimisation algorithm only included a distance limitation, and since the model did not have an explicit representation of time, it could not express nor solve dynamic problems, e.g.~including speed preferences or delivery windows. In \cite{vehits19}, we showed that the algorithm computes accurate and promising results, but has very limited expressiveness. This paper presents a substantially enhanced version of the grouping algorithm that accounts for important aspects of travel, including distance, time, speed, and cost. The model has been finalised, and simulated movement of vehicles has been introduced allowing us to model dynamic situations, where the grouping algorithm is run every time one or more vehicles or platoons meet at intersections. 

The structure of the paper is as follows; in Section \ref{sec:sota} we give an overview of the state of the art regarding platoons as well as other group building algorithms. Section \ref{sec:method} presents our proposed platoon forming and routing algorithm whose goal is finding the longest common route, while respecting, length, time, speed and cost restrictions. A simulation developed to test the approach is also presented here. Section \ref{sec:eval} presents multiple simulation experiments, which show good results regarding solution quality and algorithm runtime performance. We follow with a discussion in Section \ref{sec:discussion} where we also address the future developments of this research. Lastly we present our conclusion in Section \ref{sec:done}.

\section{Literature Review}\label{sec:sota}

The concept of platooning, where vehicles drive in a line close to each other and behaving as one unit, has been a topic of great interest in research due to it's promising outlooks for road usage on heavy-duty transports on freeways  \cite{amoozadeh2015platoon}, \cite{bengtsson2015interaction}, \cite{biswas2006vehicle}. Since vehicles travel more closely together in platoons than in usual traffic, the resource of road space is better utilised, which in turn has a positive effect on network flow. Another reason why platooning on freeways is seen as positive is the reduction in fuel consumption \cite{larson2014distributed}. Because of the reduced gap between vehicles, the wind drag does not have as high of an effect, which means that less fuel is used to maintain a constant speed. However, more recent studies and experiments have shown that although platooning did increase safety, the effect on the fuel-consumption was less than what was expected, with just a three to four percent reduction \cite{truck_platoons}


Other research on platooning deals more with control, manoeuvres (merging, splitting, travelling), stability and safety.  Ploeg et al. \cite{ploeg2013lp} calculated the needed time headway to achieve string stability and safety in platoons. This distance can be used in all scenarios, both urban and highway to calculate the spacing needed between platooning vehicles depending on speed. Another piece of research looks into cooperative manoeuvres, as an extent to platooning that again Ploeg et al. \cite{ploeg2017cooperative} studied for highway scenarios. They present a layered control architecture for such manoeuvres, which was also the basis for the Grand Cooperative Driving Challenge.

Transitioning to an urban environment, the focus still seems to be on control, stability and manoeuvres. To increase stability on curved roads and to account for the different speeds in an urban scenario, lateral and longitudinal control have to function independently according to Ali et al. \cite{ali2015urban}. A third control mechanism responsible for the platoon must also be put into place thus giving us a controller for each movement type. In \cite{khalifa2018vehicles}, Khalifa et al.~show that stability is guaranteed even with limited of communication between vehicles, through a consensus-based control mechanism for longitudinal movement. Stability of platoons is addressed in \cite{schindler2018dynamic} by making use of multiple state machines to ensure flexibility. There is one state machine for each of the logic elements of platooning: messaging, forming and distance.

Other benefits of platooning arise from a traffic management perspective in slower urban traffic. Lioris et al \cite{lioris2017platoons} have found that bottleneck-prone places, such as intersections, could benefit from platooning. By using adaptive traffic lights, and allowing platoons to pass uninterrupted, the throughput (measured in number of vehicles) of the intersection can be doubled.



Most work focusing on platoons does not touch on the aspect of how vehicles decide to platoon and who to platoon with. The study presented in \cite{schindler2018dynamic}, details just the actions needed to join or form a platoon. The aspects of how the vehicles would form a logical group and what characteristics are important in making the platooning decision lack in research. 



An interesting use-case for platoons is logistic transport, from large freight shipments, which we have already addressed, to small singular same-day deliveries. Urban platooning in the case of logistic same-day deliveries was addressed in \cite{scherr2018service} with a focus on network design. Assuming that fully autonomous platoons cannot travel on all edges of a network, human-lead platoons have to drive through those inaccessible areas. The authors goal is to minimise the costs across all vehicles in the delivery fleet of a single logistic service provider. Given the circumstance, the algorithm for vehicle grouping is not addressed, perhaps because of the large overlap in the different vehicle capabilities and objectives, as well as due to the strategic level on which the problem operates. In a similar vein, Haas and Friedrich \cite{haas2018autonomous} look at platoons in a city-logistic scenario from the traffic managers point of view. They focus on the microscopic/operational level focusing on the number of vehicles in platoons and the number of platoons in a network during the course of a single day. They consider autonomous delivery vans, that rely on platoons to travel through the network. This means that they are stationary when alone and can only drive autonomously during splitting and merging operations. They do note that the vans only join a platoon if they can be brought further to their destinations, if not they wait for another. Otherwise the authors tend to focus on developing an interaction model for platoons in a roundabout scenario, where determining who has the right-of-way is not as obvious as it is with normal traffic.


Group building algorithms have been studied for a long time and there are many variants each depending on the field of application. General approaches are presented in \cite{amir1998generic}. Other publications address specific algorithms such as \cite{wald1948optimum} for Walds SPRT Algorithm, \cite{havaldar1994extraction} which identifies important criterion and then sorts accordingly, maximum likelihood algorithms, \cite{gordon2004theories} for Wertheimers laws of grouping,  \cite{mojena1977hierarchical} for hierarchical grouping and algorithms that find the maximum consistency between the data and the group appointment.


Switching to a more specific field, namely that of traffic, we are faced again with more aspects to consider. A pre-requisite for any sort of group formation and disbandment is the ability to communicate; so Taleb et al. \cite{taleb2007stable} proposed that a group of vehicles could only exist if the communication line is maintained. This resulted in vehicles being grouped by their velocity headings; all going in the same direction, while matching the speed within the group to maintain inter-vehicular distances, and therefore the communication link. This sole criterion is not enough when formulating platoons, but does matter when considering the maintenance of one.


Ways to pair drivers with passengers while also considering schedules and routes are presented in \cite{kagaya1994use}. The problem they formulate is a many-to-many advanced reservation travel problem. Important characteristics upon which potential pairs can be made are found and studied, namely: the location of origins and destinations, the passenger type and trip purpose, closeness of desired departure times, number and capacity of available vehicles and the direction of the trips.
With normal routing algorithms the focus is on cost minimisation for the company (in the form of length of route or lack of empty trips) and not enough focus is given to the passenger and their preferences. To combat this, the authors use a fuzzy relation to find similar trips.
This work can be used as a base for our own criteria definition, although our focus is not on the match between passenger and vehicle, but rather the compatibility between vehicles.


Sanderson et al. propose a consensus based approach to clustering vehicles, in their work \cite{sanderson2012micro}. They use their own IPCon algorithm \cite{sanderson2012institutionalised} to provide collective arrangements taking preferences and constraints into account through role-associated power of the participants (vehicles in this case). Similarly, Dennisen and Müller \cite{dennisen2016iterative} propose preference-based group formation through iterative committee elections. They assume the context of a ride-sharing service that drives to different sights within a city. Passengers have to choose which sights to visit and be sorted into the autonomous vehicle based on their preference. The algorithm works by removing the unhappiest passenger using the Minisum or Minimax approval committee rules. The votes are then recounted and the process started again until a suitable solution is found. 
When considering our case, there is overlap in the fact that finding the best common route is the goal, but since we are dealing with autonomous machines and not humans, an informed vote does not seem obvious.

More related to our research field is the work of Khan and Boloni \cite{khan2005convoy} which spontaneously groups vehicles driving on freeways using a utility function based on speed. The focus is on the manoeuvres that come with convoy driving: joining, staying and splitting; but not the actions needed, but rather the utility that comes with each one. While this might be sufficient in a highway scenario, where all vehicles are driving in the same direction, it is not the case for an urban environment where vehicle routes differ greatly.

To conclude, there is only limited research activity at the intersection of the two research fields of platooning and grouping algorithms; in particular, it does not adequately address highly heterogeneous scenarios like urban logistic traffic. \pfara, described in the following section aims to address this research gap.

\section{The Platoon Forming and Routing Algorithm (\pfara)}\label{sec:method}

Based on the literature, the decisive constraints and aspects of the heterogeneous group building problem need to be determined. Taking the different elements of traffic, urban environments and logistics into account, the heterogeneity of agents and their preferences makes platoon formation and routing highly complex. First the logistic service providers will want to keep their cost minimal while reaching their target number of orders as well as keep cooperation only with some but not all other companies. The vehicles are bound by their battery life, thus restricting the route, as well as speed, be it legal limits or individual speed capabilities. And last, but most important, we have the client which selects a specific delivery window that must respected. 


The problem of heterogeneous platoon building can be solved through a linear optimisation problem using the aforementioned preferences. They are used as linear constraints in the optimisation problem that has the goal of minimising total costs. This is turn has a positive effect of both travel time and length of route. We refer to subsection \ref{opt} for a a more detailed description of the algorithm, but we do note that it is deterministic offering multiple best solutions.


Attempting to solve this problem with the aforementioned fuzzy theorem would not yield good results given the different weights given to the aspects for each vehicle or logistic service provider. For example some might prioritise battery life over speed, or some might want to wait for platoons rather than drive alone, or completely disregarding platoons if not readily available to save up on time. So while the restriction criteria are the same for all vehicles, their weight or importance will differ in the problem. 


If our algorithm fails to find a singular best solution, which could happen in networks that follow a grid-type pattern, the aforementioned voting approaches could act as fail-safe methods of selecting the grouping and route. This guarantees that the decision is not made by a singular element, but rather all potential vehicles in the platoon. 


\subsection{Input data}

The urban network is transformed into a graph to allow for routing. Intersections are transformed into vertexes and the streets into edges. Traffic demand is the sum of routes over all origin and destination points in  the network, or how many cars use each edge in a given time-span. This is meant to represent background traffic which will affect our platoon routing.
\begin{equation}\label{eq:1}
Q(x, t) = \frac{\Delta N}{\Delta t}
\end{equation} 
where \textit{Q(x,t)} is the traffic demand, $\Delta N$ is the number of vehicles and $\Delta t$ the time-span.  \cite{treiber2013trajectory}. From here we can find the traffic density, by normalising through division over the edges length. 
\begin{equation}\label{eq:2}
\rho(x,t) = \frac{\Delta N}{\Sigma_\alpha d_\alpha}
\end{equation}
where $\Sigma_\alpha d_\alpha$ is the length.

The resulting traffic density is given as each edge's weight, and act as the cost to minimise in the optimisation problem. We consider traffic density the "cost" because of its direct translation into time-savings; the freer an edge, the faster it is to transverse it, the quicker the vehicle delivers its package and can return to be dispatched again. Given the electric nature of the vehicles considered, the more time they spend away from the satellite, the longer they will need to charge, thus reducing the amounts of orders they can fulfil in a day.



Having the environment defined, vehicles can then be added and their routes calculated by \pfara. Each vehicle has an origin, a destination and a set of preferences; minimum acceptable speed for platooning, maximum speed, maximum length of route, maximum travel time and maximum cost.

\subsection{Assumptions}

The vehicles are assumed to function autonomously, all the while attempting to form a platoon. In order for a platoon to exist, a minimum of two vehicles is necessary. They must be at the same vertex at the same time to do so. For now vehicles do not wait at intersections for other vehicles or platoons, even if their preferences allow it. Platoons form spontaneously and organically with vehicles that would benefit from platooning. A vehicle would not participate in a platoon if the cost to do so is larger than travelling alone, or its distance, time, speed and cost restrictions are not respected. After a platoon forms, all vehicles in it drive with a uniform speed. 
As pointed out in the literature review, vehicles must have communication capabilities, based on \cite{taleb2007stable} and \cite{khalifa2018vehicles}. Each vehicle sends its destination and restrictions to a local agent which performs \pfara. The results (route, expected cost, expected length and group) are then communicated back to each vehicle. 

\begin{figure}
\includegraphics[width=0.75\textwidth]{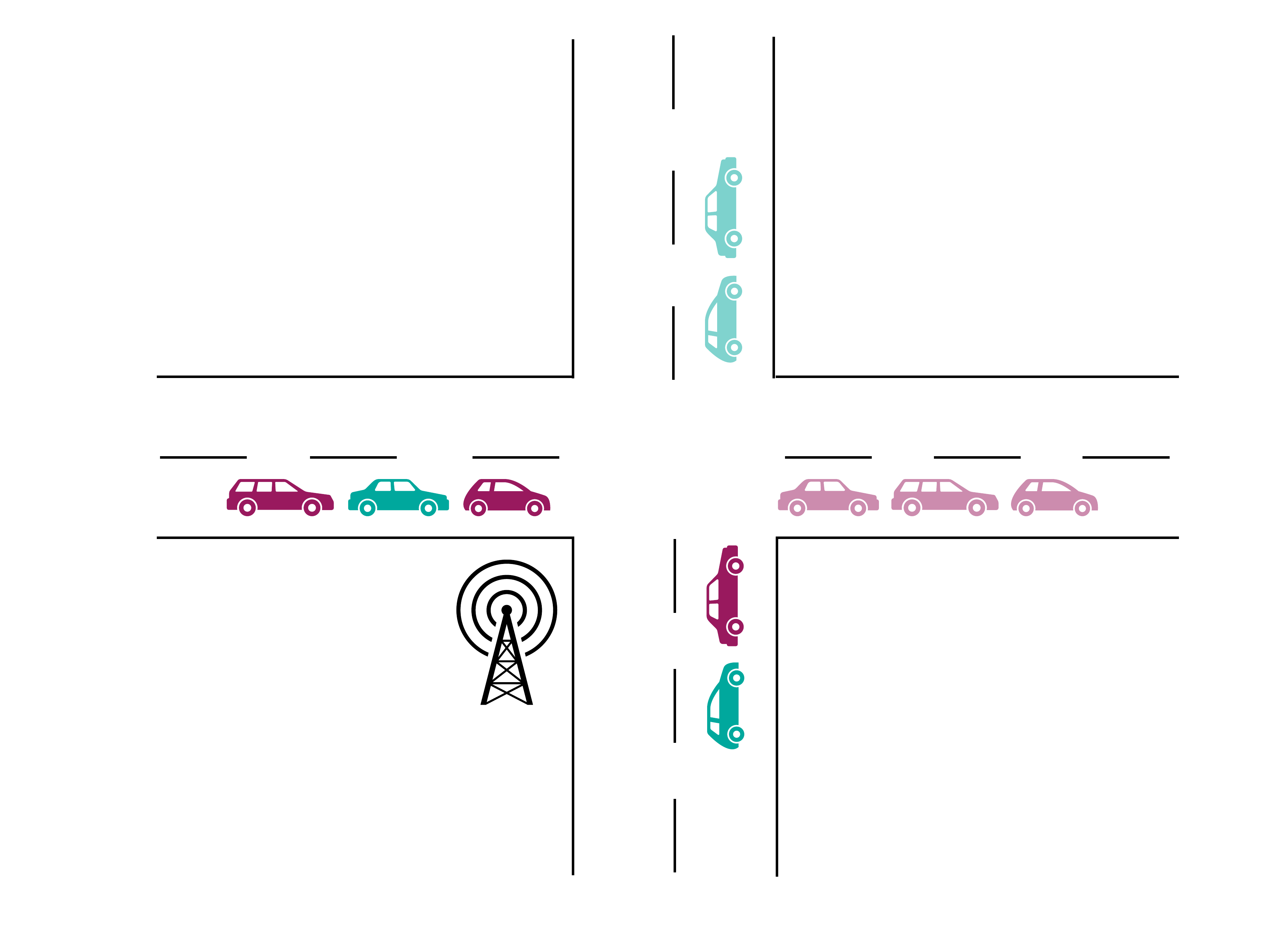}\centering
\caption{Vehicle interaction and group formation.} \label{fig:intersection}
\end{figure}

For a visual representation of how platoons would form, please see Figure \ref{fig:intersection}. Two platoons approach the same intersection, but the vehicles comprising them follow different headings, denoted by the different opaque colours. The vehicles communicate their destination, preferences and limitations to the local agent denoted by the broadcast tower, which runs \pfara and forms two new platoons, denoted by the slightly more transparent vehicle formations.

To encourage the penetration rate of platoons, they are treated as one vehicle, since logically they perform as one. All members of the platoon share a common route and contribute equally to its cost. This leads to a reduction of individual costs for all vehicles and encourages the formation of larger platoons. The cost of a route is considered as the sum of weights for all edges used in it.





\subsection{Notations}

For ease of understanding, all symbols to be used in the formal description of the algorithm and their meaning is presented in Table \ref{tab:defs}.
\begin{table}
\caption{Symbols and meaning \cite{vehits19}.}\label{tab:defs} \centering
\begin{tabular}{ | c | c | }
\hline
Symbol & Definitions \\
\hline
G=(V, E) & graph\\
$n \in$ V & vertices\\
$e \in$ E & edges, $e \equiv (i, j)$\\
$d(e)$ & traffic density of edge $e$\\
$l(e)$ & length of edge $e$\\
$s$ & speed of the platoon \\
$v$ & vehicle \\
$dest_v$ & destination vertex of vehicle $v$ , fixed\\
$K^{*}_v$ & maximum costs for vehicle $v$\\
$\Omega_v$ & maximum delivery time for vehicle $v$\\
$\Lambda_v$ & maximum length of route for vehicle $v$\\
$p(e)_{v}$ & price vehicle $v$ pays for edge $e$\\
$NP$ & the number of vehicles in the platoon\\
\hline
\end{tabular}
\end{table}

\subsection{\pfara: The Speed-Clustering Step }

Given that all vehicles in a platoon must travel with the same speed to maintain stability, an initial speed clustering has to be put in place. A minimum acceptable speed for platoons as well as the maximum speed is specified in the vehicles preferences. To form a grouping based on speed, the maximum minimum speed within the group is first selected. All vehicles whose maximum speed is greater than it are grouped and their destination and preferences directed to the second step: the optimiser. If not all vehicles in the original group were selected, the next greatest minimum speed is taken, and the same procedure applied, to be repeated until all vehicles are grouped, or it is determined that they must travel alone. This can be written as:
\begin{lstlisting}
while not (all vehicles grouped or excluded)
{ 
 find maximum minimum speed;
 for each vehicle 
 {
   if (maxspeed >= maxminspeed) 
      group vehicle;
 }
}
\end{lstlisting}

While determination of the platoon's speed could be written as a linear constraint in the following step, we believe that this approach would be restrictive. The common speed calculated would be lower than an ideal speed and could possibly deter vehicles from joining. This means that our algorithm has to be repeated with a smaller group of vehicles which not only takes time, but also provides less cost savings.

\subsection{\pfara: The Optimiser Step}\label{opt}

The objective for this step is to find the route with the most overlap between vehicles that also respects their restrictions. This can be achieved by adapting the shortest path algorithm (presented by equations \ref{eq:3},\ref{eq:4} and \ref{eq:5}) to fit multiple vehicles instead of one.

\begin{gather}
min \sum_{(i,j) \in E} x(i,j)*d(i,j) \label{eq:3}\\
x(i, j) \in {0,1}  \forall edge\ (i,j) \label{eq:4} \\
\sum_j x(i,j) - \sum_j x(j,i) =  \begin{cases}
1\ \iff \ i =Origin \\
-1\ \iff \ i = Destination\\
0\ otherwise.
\end{cases} \label{eq:5}
\end{gather}
$x$ is the variable associated with an edge, defined between vertices $i$ and $j$. It has the value 0 if its edge is not in the shortest path, or 1 if it is (\ref{eq:4}). The flow constraint enables the vehicles to pass through intermediary nodes and is defined with the three cases (\ref{eq:5}). Naturally the objective is to minimise the costs which we measure in traffic density (\ref{eq:3}).

When transposing this to platooning, we introduce a new variable  $y$ to signify the platoon. This variable acts in a similar fashion to how $x$ did in the classic problem: $y= 0$ if it is not part of the route of any vehicle and $y=1$ if it is (\ref{eq:7}). $y$ is used in the objective function to be minimised (\ref{eq:6}). To ensure that the route covers all vehicles, we add a restriction that states that $y$ can have the value of one only if there is at least an $x$ with the value of one (\ref{eq:8}). The flow constraint remains unchanged (\ref{eq:9}). To these restrictions we add the ones for length of route (\ref{eq:10}), delivery time (\ref{eq:11}), and maximum cost (\ref{eq:14}). Incentivising platoons is done by giving a reduction in costs. Since all vehicles contribute to the cost (\ref{eq:13}), the more vehicles in a platoon, the less they have to pay for each edge.
Therefore the platooning optimisation problem would be formulated as such follows:
\begin{definition}
Given a group of vehicles, being at the same vertex $O$ in the graph at the same time; let the routes for all the vehicles be given by $y(i,j)$, and the individual vehicle routes by $x(i,j)_v$ where
\begin{gather}
min \sum_{(i,j) \in E} y(i,j)*d(i,j) \label{eq:6}\\
y(i,j), x(i,j) \in \{0,1\},  \forall edge\ (i,j) \label{eq:7}\\
x(i,j)_v \leq y(i,j) \forall edge\ (i,j), \forall vehicle\ v \label{eq:8}\\
\sum_j x(i,j)_v - \sum_j x(j,i)_v =  \begin{cases}
1\ \iff \ i =O \\
-1\ \iff \ i = dest_{v}\\
0\ otherwise.
\end{cases}
\forall v,\ \forall (i,j) \in E \label{eq:9}\\
\sum_{x(i,j)_v=1} l(i,j) \leq \Lambda_{v} \forall\ v \label{eq:10}\\
\sum_{x(i,j)_v=1} \frac{l(i,j)}{s} \leq \Omega_{v} \forall\  v \label{eq:11}\\
p(i,j)_v \geq 0  \ \forall \ v\ and\ edge\ (i,j) \label{eq:12}\\ 
p(i,j)_v = \frac{d(i,j)}{NP_{(i,j)}} \forall \ v \ where\ NP_{(i,j)} = \sum_v x(i,j)_v \label{eq:13}\\
\sum_{(i,j):x(i,j)_v=1}p(i,j)_v \leq K^{*}_{v}, \forall\ v . \label{eq:14}
\end{gather}
\end{definition}

A similar but much less expressive model was presented in \cite{vehits19}. It only addressed the optimisation problem limited strictly by a length of route constraint. Also the analysis of the grouping algorithm was done statically since the model did not feature time, and by extension, speed. Without those elements a complete solution accounting for the four most important aspects of travel (length, time, speed and cost) was not possible. 

This work features all the aforementioned aspects and and takes into account the time of meeting, speed matching and the cost restrictions. This ensures the solution provided is complete and can be applied to all vehicles regardless. 

\subsection{Simulation}

The simulation was formulated to study the effectiveness of \pfara  and its effects on the valuation of cooperation between vehicles.  Its performance is also tested to ensure that it is applicable and appropriate for real-life scenarios.

\subsubsection{Framework}
To analyse \pfara, a simulation was designed using Java. Each separate component has a specific framework.
Jung \cite{jung} is used to generate the environment due to its powerful library that models data into a graph, its visualisation options, routing and analysis capabilities. To visualise the routes and traffic densities, we employed the use of a heat-map using the colour schemes provided by \cite{colormap}. The actual drawing of routes was done with a JXMapViewer2 painter, modelled after a pre-existing examples \cite{jxmap}.
For the optimisation part, Gurobi \cite{gurobi} was used, a powerful commercial solver. It was extremely easy to implement into the simulation as an external jar file and the problem definition process was also simple. To write the problem we start by defining as many $x$ variables as we have vehicles in the potential platoon, and one $y$ variable. Then the objective function is written based on the $y$ and weights (traffic density) of the edges with minimisation being the goal. Then for each of the criteria defined above (equations 8, 9, 10, 11 and 14), a linear constraint is written. Since speed is already addressed before running the optimiser, we do not need to include it here. After running and finding a solution, the results are saved in a separate data structure, available at any time and the optimiser object made redundant to save up on computational resources.

\subsubsection{Methodology}
Simulation starts with creating the environment and the vehicles from an input file. The vehicles each have their set of preferences as well as a starting satellite and a destination. Then the internal clock starts and after running instances of the optimiser for each satellite to group the vehicles, they begin moving through the environment. They go along their given route, attempting to form new platoons whenever they encounter new vehicles/platoons. If a vehicle or a platoon intersects with other vehicles or platoons, they go through the grouping process (first by speed and then by optimiser). The algorithm finds the best group, their routes (both common and separate until their destination) and expected cost for each vehicle.
After grouping, the vehicles travel together until the end of their common route, where they split and continue their trip towards their destination, or attempt to form a new platoon with some of their former co-platooners. To ensure that the selected route is constantly the "best" one, given that urban traffic tends to shift quite rapidly, \pfara can be performed at some or all intermediary nodes.
To account for the distance and time travelled as well as the cost of the route, and update of the preferences is performed at each intermediary node. The length of the edge travelled is subtracted from the maximum distance, and the same goes for the maximum time and cost. This ensures that a new instance of the optimisation part of the algorithm sets the restrictions to a current version of the preferences, and not to the ones at the start of travel.
 The simulation completes with the vehicles reaching their respective destinations. Afterwards, an output file is written detailing each action of each vehicle, their cost, length of route and time of arrival. Lastly a heat map of the network is generated allowing us to see which edges were used by the vehicles and of how many of them were on each edge.

\subsubsection{Baseline algorithm: Overlapping}
To have a benchmark comparison for our approach we decided to conceptualise the most obvious version of a platooning grouping algorithm, which is simple best route matching. Considering each vehicles current position as the origin point, the already-existing routing algorithm present in Jung is used to find the fastest route. By aggregating the routes for all vehicles, we can find where they would travel together (considering constant speed). By counting the number of times an edge was used, we determine the potential size of the platoon travelling on it as well as knowing which vehicles make up the platoon. This approach however does not take into account the possible restrictions, re-routings and above all, vehicle preferences. It is on the other side extremely easy to calculate and straightforward. To ensure we are just comparing the optimisation part to this simple overlapping approach, all other aspect of the simulation were used in the overlapping algorithm as well. Therefore we have the time counter, the vehicles' movement throughout the environment, and most importantly, the preliminary speed clustering.


\subsubsection{Input Data}

To simulate scenarios we employed data from \cite{transportationnetworks}, namely the Berlin Tiergarten neighbourhood. The data set is finely granular enough to use and comes  with a very rough account of traffic demand. It was not given for each edge in particular, but rather as an aggregation of trips based on zones in the neighbourhood. So to get accurate weights for our problem, individual trips needed to be generated. The vertices were divided into zones and random ones pulled to act as origins and destinations. Routes were calculated, aggregated for the whole instance, and then normalised by dividing through each edges length (according to the definition given by Equation \ref{eq:2}), thus giving us traffic density to act as the edges weights for our algorithm. An example of this process would be having thirteen trips from zone 2 to zone 9, pulling a random node from zone 2 and one from zone 9, finding the best route between them, saving the results and repeating the process another twelve times aggregating the edges taken for all. This process does not offer constant results, but they are similar enough between runs to be considered consistent. 

We also created a synthetic smaller network of a five-by-five grid network to follow the routes and examine the driving behaviour more closely. Factors like traffic density, positioning of satellites and vehicle preferences were varied to ensure the generality of \pfara.



\subsubsection{Output Data}

As mentioned before, after each run the simulation provides output in the form of a log file and a heat map visualisation. The colours selected to display it (Inferno palette) were chosen due to the cognitive ease of understanding it by the viewer \cite{thyng2016true} (yellow for hot to dark purple for cold). A light yellow edge means it was the most used, whereas dark purple indicates it was a part of the route of a single vehicle. A colourless edge signifies it not being a part of any vehicles' route.


The log file consists of the events of all vehicles. The time step and location of each event is specified. The event types are: 
\begin{enumerate}
	\item Creation. Takes place at the satellite vertex.
	\item Departure. Happens for satellite and intermediary vertices.
	\item Arrival. Happens for intermediary vertices as well as the destination.
	\item Completed. Accompanies the arrival event to denote the completion of the route.
	\item Formed. When a vehicle joins a platoon and who they join with.
	\item Split. When a vehicle disbands from the platoon. Can be due to completion or because the end of the common route was reached.
\end{enumerate}

Besides the events, upon completion the following is specified:
\begin{enumerate}
	\item the cost accumulated
	\item length of route taken
	\item the allowance left for length, time and cost for all vehicles.
\end{enumerate} 

The overlapping algorithm provides the same result structure; a heat map and a log file.

\section{Results}\label{sec:eval}
\noindent
To check the validity of our algorithm, we ran experiments on real and synthetic environments. The real environment was the Berlin Tiergarten neighbourhood that also has realistic traffic density. The synthetic one is represented by a five by five Manhattan grid network. To validate all aspects of the algorithm, multiple experiments were run varying things like traffic density, satellite placements, and vehicle preferences. The algorithm did well in all cases and all groupings and routes were correct and respected the restrictions. 

By looking at the heat maps, we get an idea of how the placement of satellites and traffic density affect the deliveries, and where joining and splitting could occur across the platoons. The colours show how many vehicles use a specific edge and a change in colour from one edge do the next may denote vehicle/s splitting (if it shifts colder) or joining (if it shifts warmer). However, due to fact that the heat map is generated statically at the end of the simulation, it cannot serve as a definite way of determining said points. The colour of the edge changes with how many vehicles use it and not necessarily just with platoons. However we will present a case where the split and join points can be clearly identified.  

\begin{figure}
\includegraphics[width=\textwidth]{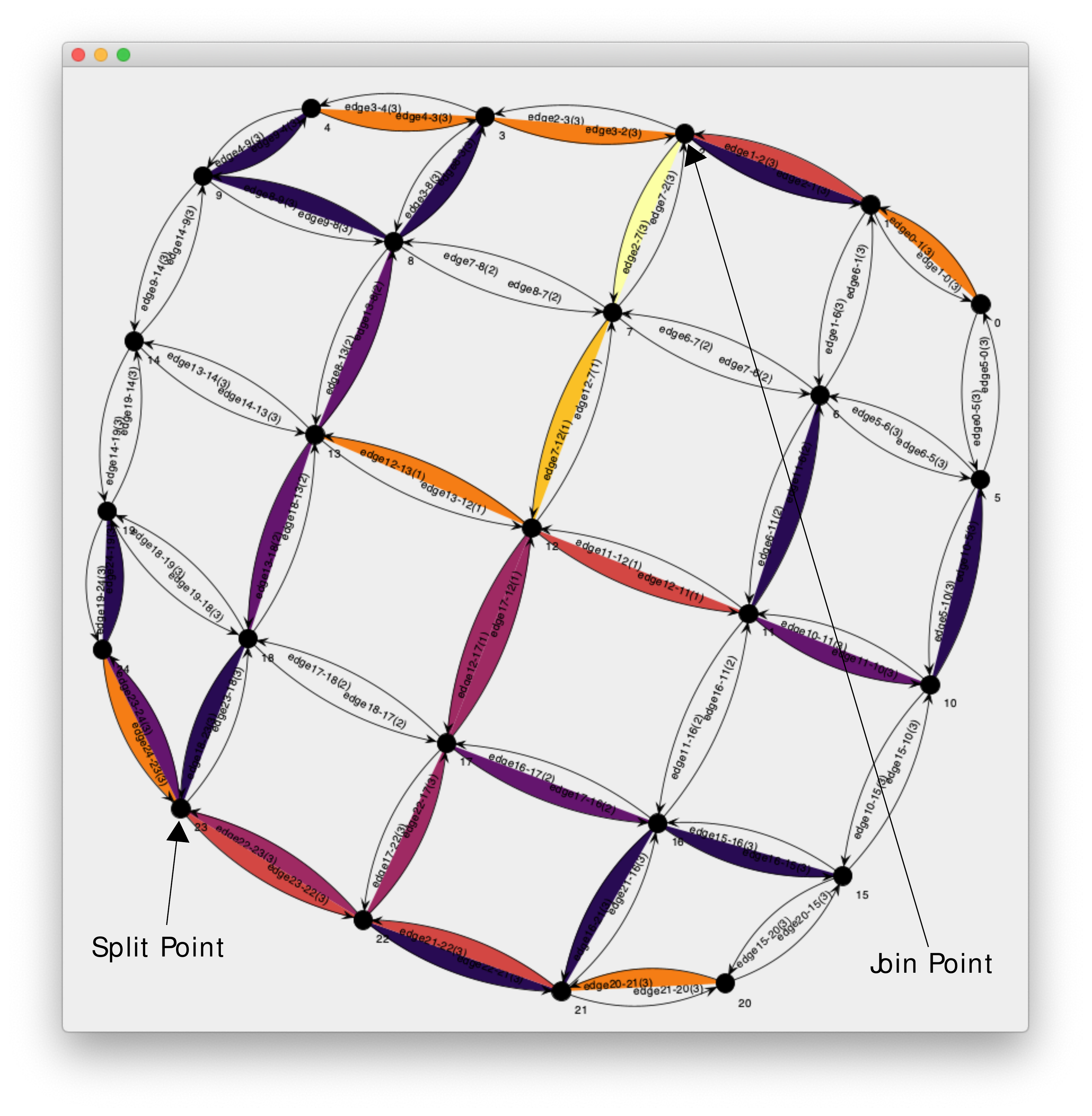}\centering
\caption{Heat map depicting split and join points.} \label{LtoHIO}\centering
\end{figure}

We take the example of the five by five Manhattan grid network, with the outside edges having higher traffic density, which decreases towards the centre. We simulate five different logistic service providers, each with a fleet of five vehicles, each having a random destination node. The locations of the satellites are the four corners of the network and the central node. All the vehicles start at the same time (the starting time of the simulation) and travel at the same speed. Split points can be identified by a transition of the edge colour from warm to cooler when looking at Figure \ref{LtoHIO}. That is the case at node 23 where a vehicle splits from the five vehicle platoon coming from node 24 and continues alone towards node 18, with the other four continuing their journey in the platoon towards node 22. Joining points like node 2 can be identified by the transition of the edge colour from cooler to warmer. Two platoons, one coming from node 3 containing five vehicles, and the other coming from node 1 containing four vehicles, meet and join at node 2 in a larger seven vehicle platoon (node 2 was the destination for two vehicles) travelling toward node 7.



\subsection{Numerical results}
To analyse the numerical results as best as possible we compared our approach to independent travel and the overlapping algorithm. Pinning the two grouping algorithms agains one another allows us to see how viable our design is. The real environment of Berlin Tiergarten Neighbourhood was used to guarantee the results are as realistic as possible. To resemble real traffic we assumed the satellites are located at intersections where most of the traffic takes place (number of vehicles leaving the intersection is highest) and the vehicle destinations we selected the nodes where the number of vehicles arriving would be highest, both based on traffic density information. We have two providers, each with one satellite and 15 and 10 vehicles respectively, each with a different destination. With the origin and destination points set, both algorithms were run and the resulting costs analysed.

\begin{figure}
\includegraphics[width=0.8\textwidth]{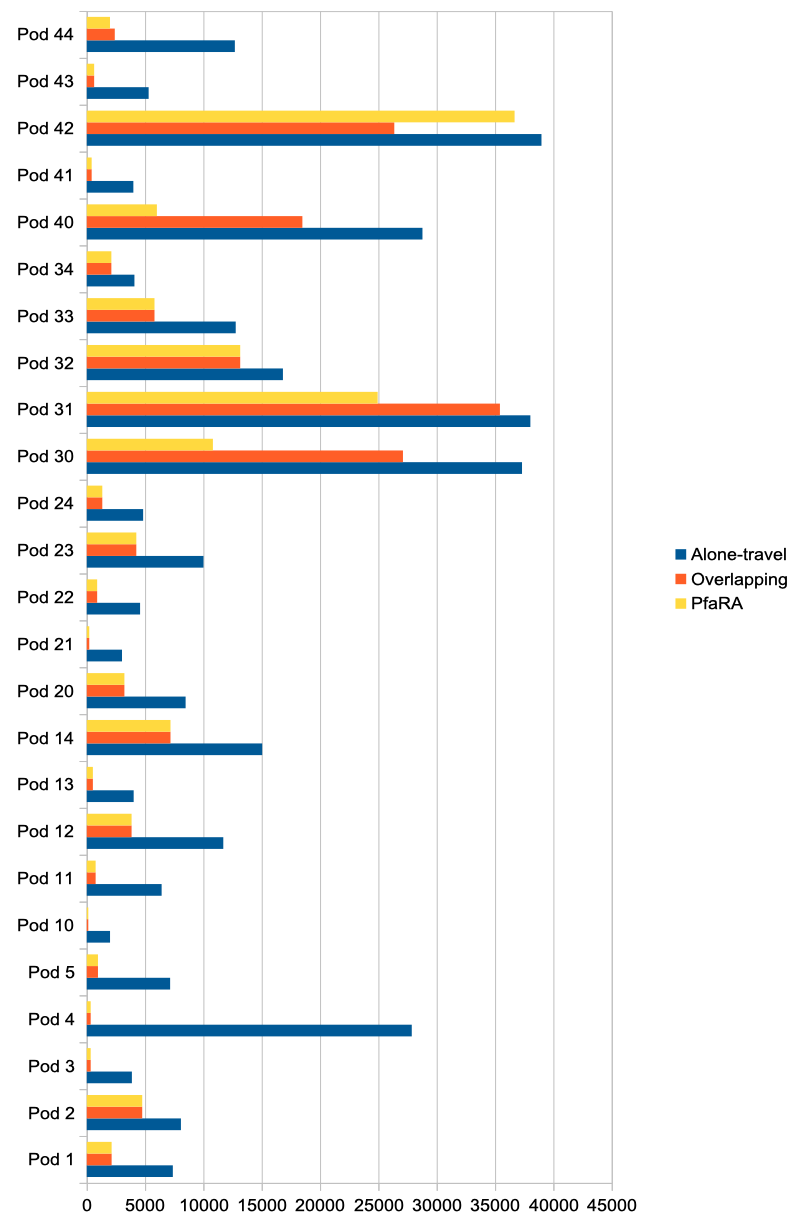}\centering
\caption{Cost comparison of the two algorithms with alone-travel on realistic network.}\label{fig:plot1}
\end{figure}

From 
Figure \ref{fig:plot1} we can see what sort of cost savings we accrue from using the two grouping algorithms. In the case of the first provider (pod 1 to pod 24), both the overlapping and the \pfara approach mostly give the same results. The only two exceptions are pod 13 and 24, which have less costs with \pfara. This shows us that this approach found an alternative route that allows for longer platooning for at least one of the vehicles, but which benefits both. This is confirmed when analysing output log file.

In the case of the second provider (pod 30 to pod 44) we have more variety in the results. Again, both algorithms are far superior than the alone-travel alternative, but when comparing them to each other we have an array of relations. In some cases, like pod 43, pod 32, the cost is the same. Pod 40 and pod 30 have a significantly better result with \pfara. The most interesting result is pod 42, whose costs for both grouping algorithms are lower than travelling alone, but \pfara does considerably worse than the overlapping alternative. This means the route given by \pfara for this vehicle is longer, but when considering the global cost of this provider, it might be worth it. The total costs for this logistic service provider are:
\begin{itemize}
\item Alone: 198761
\item Overlap: 121897.97
\item Optimiser: 102558.95
\end{itemize}

We can see that even if some vehicles take detours, it might be worth for the service provider to go for the \pfara approach anyway. The influences and weights of each of these factors (cost, length and time) need to be calculated and specified by each of the service providers, and adjusting the vehicles preferences and restrictions accordingly. Table \ref{tab:comp} shows a more detailed view of the results for the specific cases mentioned above.

\begin{table*}[ht]
\caption{Result comparison of the two algorithms, detailed.}\label{tab:comp} \centering
\begin{tabular}{ | c | c | c | c | c | c | c | }
\hline
Vehicle & $Cost_{Overlap}$ & $Time_{Overlap}$ & $Length_{Overlap}$ & $Cost_{\pfara}$ & $Time_{\pfara}$ & $Length_{\pfara}$\\
\hline
pod 43 &633.87 & 45 & 3.74 & 633.87 & 45 & 3.74\\
pod 32 & 13154.73 & 27 & 1.29 & 13164.73 & 27 & 1.29 \\
\hline
pod 40 & 18482.71 & 81 & 6.01 & 6013.36 & 81 & 6.01\\
pod 30 & 27100.71 & 101 & 6.85 & 10816.02 & 101 & 7.44\\
\hline
pod 42 & 26364.71 & 93 & 7.40 & 36673 & 107 & 8.09\\ 
\hline
\end{tabular}
\end{table*}

Another experiment is ran on the synthetic five-by-five grid network to present an easier to understand set of results, which is not possible with the Tiergarten neighbourhood due to its size. In this case there are five logistic service providers each with five vehicles.


\begin{figure}
\includegraphics[width=0.8\textwidth]{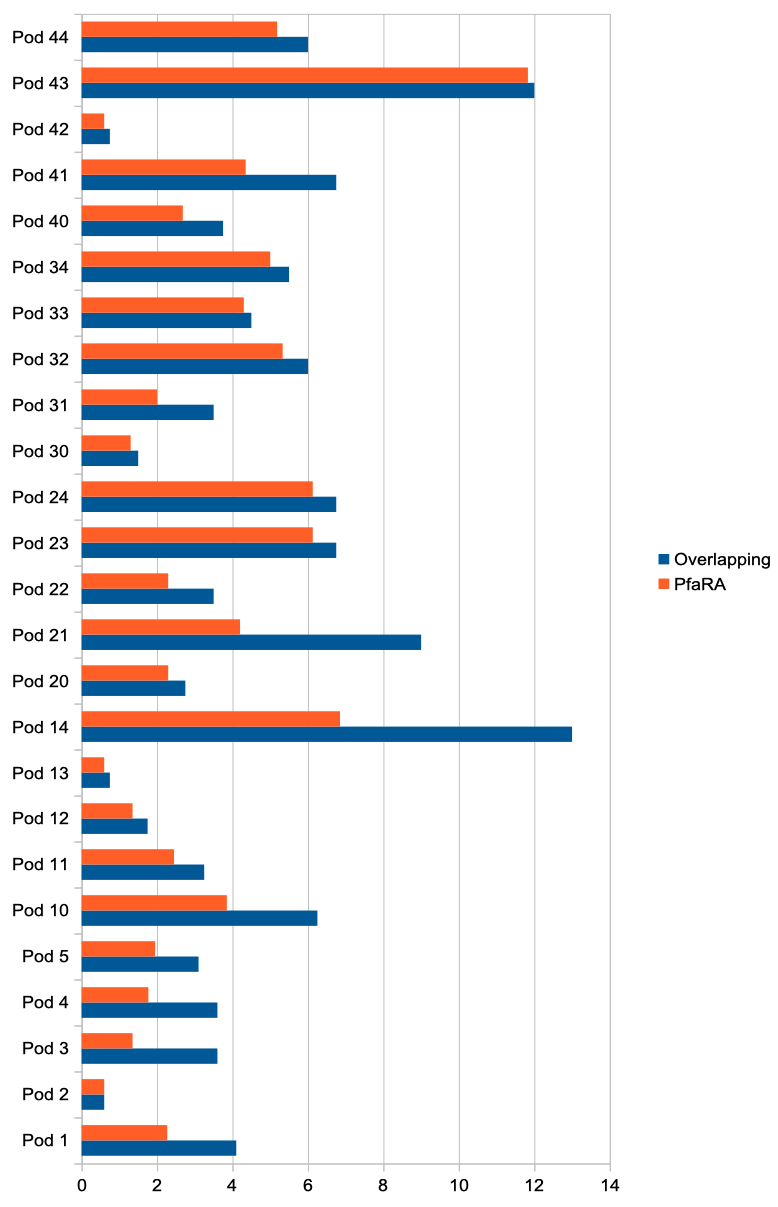}\centering
\caption{Cost comparison of the two algorithms on synthetic network.}\label{fig:plot2}
\end{figure}

We can tell from 
Figure \ref{fig:plot2} that the optimiser approach provides a better result than the overlapping approach for all vehicles. Usually the vehicles do not stray from their best route, which is shown by observing both the length of route and travel time given in the output file. The only exception is vehicle 43, which takes a rather considerable detour to travel with a platoon. Even so, the cost it accrues is still lower than the alone travel alternative, which is 12. One could argue that this cost saving is not worth the extra time and distance traveled, since the vehicle could have returned to the satellite and been re-dispatched on another delivery. As mentioned before the cost-time-length valuation is up to each logistic service provider to decide and implement it through the vehicles preferences.

When looking at the results for both scenarios, we can see that sometimes the simplistic overlapping approach does offer better results than the optimiser, but globally it is not the best solution. The optimiser guarantees us the group specific best solution, so the minimal cost for all vehicles considered, leading to a Pareto-type solution. Another fault of the overlapping algorithm is its simplicity, none of the preferences and limitations imposed on the vehicles are considered, leading to a possible incomplete solution. The optimiser takes limitations and preferences into account, giving not the universal (impossible to calculate on a large-scale scenario), nor the user (vehicle) best solution, but somewhere in the middle, with a group specific best.




Besides the improvement in cost, we also tested the influence that the preferences of the vehicle have in \pfara. Restrictions in speed, length of route and time of travel affect the grouping of vehicles in the form of the preliminary clustering (speed) as well as in the optimisation problem (time and length). For that we modified the scenario presented previously, and made vehicle 5 slower, vehicle 41 to have a length limitation and vehicle 43 to have a time limitation. The results are presented in Table \ref{tab:mod} having the first three columns with the cost, time and length of the route for the \pfara without restrictions and then the last three columns he cost, time and length of the route for the \pfara with restrictions.

\begin{table*}[ht]
\caption{Comparison of \pfara on synthetic network, with and without different vehicular preferences.}\label{tab:mod} \centering
\begin{tabular}{ | c | c | c | c | c | c | c | }
\hline
Vehicle & $Cost$ & $Time$ & $Length$ & $Cost_{Pref}$ & $Time_{Pref}$ & $Length_{Pref}$\\
\hline
pod1 & 2.27 & 55 & 500 & 2.783 & 55 & 500\\
pod2 & 0.6 & 11 & 100 & 0.75 & 11 & 100 \\
pod3 & 1.35 & 22 & 200 & 1.75 & 22 & 200\\
pod4 & 1.77 & 33 & 300 & 2.25 & 33 & 300\\
\rowcolor{gray}
pod5 & 1.945 & 44 & 400 & 9 & 67 & 400 \\
\hline
pod10  & 3.85 & 44 & 400 & 3.85 & 44 &400 \\
pod11  & 2.35 & 33 & 300 & 2.35 & 33 & 300\\
pod12 & 1.35 & 22 & 200 & 1.35 & 22 & 200\\
pod13 & 0.6 & 11 & 100 & 0.6 & 11 & 100\\
pod 14 & 6.85 & 55 & 500 & 6.85 & 55 & 500\\
\hline
pod20 & 2.295& 55 & 500 & 2.4 & 55 & 500 \\
pod21 & 4.2 & 33 & 300 & 4.2 & 33 & 300 \\
pod22 & 2.295 & 55 & 500 & 2.9 & 55 & 500 \\
pod23 & 6.128 & 77 & 700 & 6.23 & 77 & 700 \\
pod24 & 6.128 & 77 & 700 & 6.23 & 77 & 700 \\
\hline
pod30  & 1.3 & 22 & 200 & 1.3 & 22 & 200\\
pod31 & 2 & 22 & 200 & 2 & 22 & 200 \\
pod32 & 5.3 & 33 & 300 & 5.3 & 33 & 300 \\
pod33 & 4.3 & 33 & 300 & 4.3 & 33 & 300 \\
pod34 & 5 & 33 & 300 & 5 & 33 & 300 \\
\hline
pod40 & 2.683 & 44 & 400 & 4 & 44 & 400 \\
\rowcolor{gray}
pod41 & 4.35 & 33 & 300 & 7.5 & 33 & 300 \\
pod42 & 0.6 & 11 & 100 & 1.5 & 11 & 100 \\
\rowcolor{gray}
pod43 & 11.183 & 88 & 800 & 8 & 44 & 400 \\
pod44 & 5.183 & 66 & 600 & 6.5 & 66 & 600 \\
\hline
\end{tabular}
\end{table*}

We can see that the vehicles 30 to 34 and 10 to 14 are not affected by the preference modifications. Vehicles 1 through 4 are affected by the lack of vehicle 5 in the platoon,  it being excluded due to its lower speed. This is turn also slightly affects the cost of vehicles 20, 22, 23 and 24. Given vehicle's 43 time and 41's length restrictions the previous formation and route of that platoon completely changes. Instead of a five vehicle platoon, we have two platoons, of two and three vehicles respectively. 

\subsection{Performance}
\begin{figure}
\includegraphics[width=\textwidth]{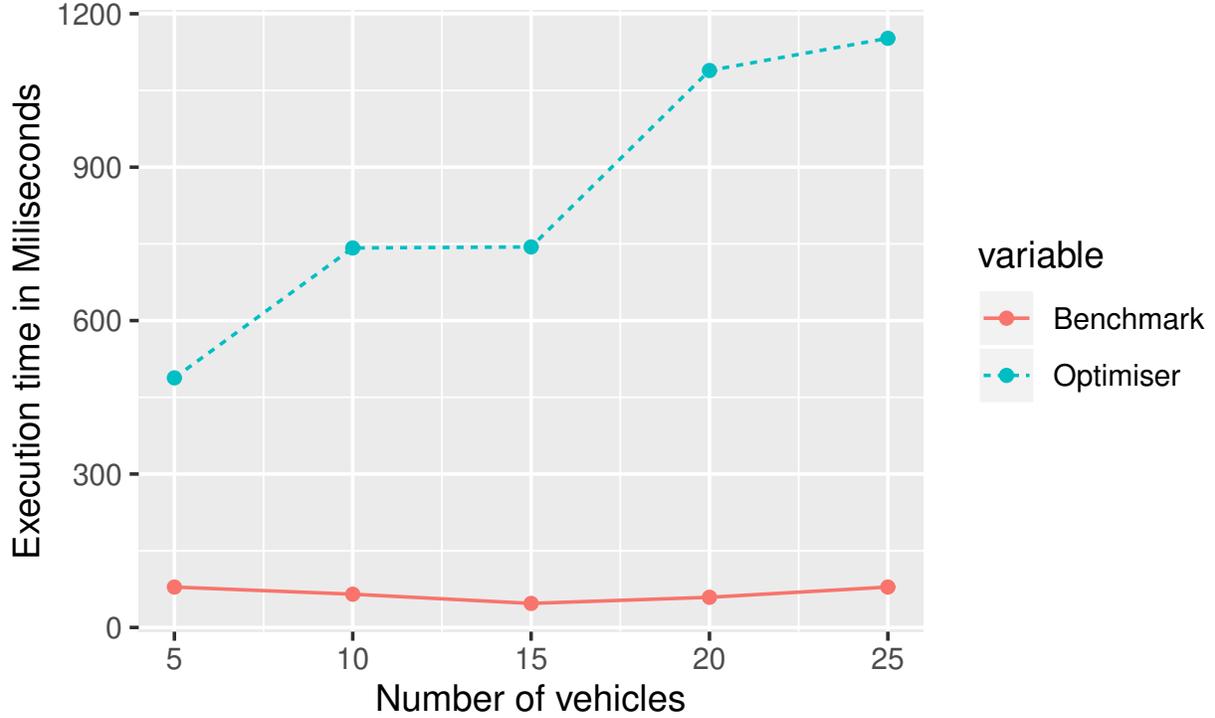}\centering
\caption{Performance analysis.}\label{fig:perf}
\end{figure}

Figure \ref{fig:perf} shows both algorithms' performance, representing the runtime of the \pfara and the overlapping algorithm respectively for the number of vehicles specified. These experiments were run on the realistic environment of the Berlin neighbourhood to guarantee \pfara could be applicable in a real-life scenario. When looking at the runtime of the algorithms, an evident increase happens with raising the number of vehicles. Even so, with 25 vehicles being considered for platooning at the same time, a result is still found in under two seconds. Considering that vehicles must not only be physically near each other to platoon, but also close time-wise, we can safely assume that in a real traffic scenario, no more than 20 vehicles would attempt platooning. In the case where a large platoon would meet another, due to them logically being one vehicle, \pfara is also easily solvable. With the more likely case of having five or ten vehicles, \pfara does really well, finding the best solution in 0.5 and 0.75 seconds respectively. 

\section{Discussion and Future Work}\label{sec:discussion}

The algorithm presented in this paper provides a viable solution to the platoon forming and routing problem, for vehicles that are heterogeneous in nature and have limited capabilities.

When comparing our approach to the state of the art, \pfara takes into account multiple aspects aspect of travel (speed, time, length and cost). It does slightly underperform when comparing it to a more straightforward route-matching approach, but respects individual preferences and restrictions of vehicles and hence allows us to consider cross-provider platooning. To the best of our knowledge there are no other approaches to group vehicles in a platoon from a logical standpoint; as mentioned before, focus tends to be on the manoeuvres necessary to form, rather than what makes the vehicles form.

When comparing to the results in our previous work, having the same input data, we can see that the inclusion of time, speed and movement has a significant influence on the costs. More limitations on the system means less flexible grouping; but this also means that the groups are robust and a spontaneous change of route that could lead to disbandment is less likely. The new additions to the model have not affected the runtime of the algorithm negatively; on the contrary, PFaRA, with its preliminary speed clustering and enhanced optimisation problem performs faster than its predecessor presented in \cite{vehits19}. This is due to the added restrictions which discards non-plausible routes faster, significantly reducing computational time.

\pfara performs well, giving solutions for platooning for all vehicles considered; whether they start at the same satellite or meet along the way. The routes provided are the group specific best solutions, thus accounting for heterogeneity of the vehicles, providing each vehicle with cost savings. A user specific optimum would most likely exclude that user from the group because it deviates from the Pareto solution provided. A global best solution would be hard to calculate, due to the size of the network, and would most likely not be viable long term (for example for the entire length of the route) due to the fact that urban traffic is considerably volatile. This is why we focus on finding a group specific best solution and due to the structure of the algorithm, the system can be relaxed or restricted by removing or adding in constraints.

Our approach can be applied to arbitrary platoons and is not restricted to just logistic traffic. It is general enough for all types of vehicles and environments given that the vehicles preferences are known. 
One limitation of our current approach is that it does not consider the physical size of the vehicles. In urban traffic space is limited, and while heterogeneous platoons also include differently sized vehicles, one must guarantee that there is enough space to accommodate it. This is where the time headway proposed in \cite{ploeg2013lp} can be used to calculate the physical size of the platoon based on it's speed and the sizes of the vehicles included in it. We are planning to address this aspect in future research. 


A further important extension of our model will include adding a negotiation algorithm that allows for vehicles to arrange further travel in a platoon through monetary exchanges at the end of their common route. This means that any vehicle accepting will be taking a detour from their ideal (selfish) route. The trade-off between cost reduction and travel limitations is up to each vehicle to calculate. Moreover due to the competitive nature of the LSPs, the compensations offered need to be high enough to be profitable for competitors in order to convince them to participate, while also being low enough for the offerer to maintain an upper hand in the market. Problems usually associated with this type of scenario such as deliberately accepting less than favourable routes to financially "hurt" the competition which also means waste of resources should be avoided.

\section{Conclusion}\label{sec:done}

The use of autonomous electric vehicles for last-mile deliveries in urban areas can bring benefits to customers, traffic managers and logistic service providers alike. They no not create emissions, can navigate through traffic alone and remove the need of a human driver. But the advantages they bring can be increased by having them travel in platoons when possible.


To guarantee the formation of platoons, all delivery vehicles from different logistic service providers should be willing to cooperate, thus creating heterogeneous groups. The vehicles have preferences (the minimum speed they are willing to accept for the platoon), restrictions (the end of the delivery time window, or travel distance manageable with their battery autonomy) and characteristics (maximum speed they can achieve) that have to be taken into account.


Our contribution is a grouping and routing algorithm for heterogeneous vehicles with same-day delivery in mind. The vehicles go through two grouping stages; one based on speed and another based on route and restrictions. The second stage is an optimisation problem that looks for the largest route overlap between the vehicles, while respecting their limitations (in the form on linear constraints). The objective is to minimise the cost of the vehicles considered, which is measured in traffic density. This approach offers feasible groups and routes, while maintaining the possibility of further relaxation or limitation of the system.


The  proposed \pfara performs well in terms of solution quality (cost) and runtime performance, offering a viable and robust grouping and routing for all vehicles, independent of when and where they meet.


\section{Acknowledgements}
\noindent This work has been funded by the Deutsche Forschungsgemeinschaft (DFG, German Research Foundation) under Grant 227198829 / GRK1931. The focus of the SocialCars Research Training Group is on significantly improving the city‘s future road traffic, through cooperative approaches. We also acknowledge Nelly Nyeck for her help with gathering results, Philipp Kraus for suggesting frameworks and Stephan Westphal for guidance towards a correct optimisation formulation.
\bibliographystyle{unsrt}
\bibliography{SebePFaRA.bib}

\end{document}